\documentclass[authoryear,preprint,review,12pt]{elsarticle}

\usepackage[a4paper,margin=2cm]{geometry}
\usepackage{amsmath,amssymb}
\usepackage{graphicx}
\usepackage{hyperref}
\usepackage{booktabs}
\usepackage{subcaption}
\usepackage{algorithm}
\usepackage{algorithmic}
\usepackage{todonotes}

\hypersetup{colorlinks=true,linkcolor=black,citecolor=black,urlcolor=cyan}

\journal{Computer Physics Communications}

\begin{document}

\begin{frontmatter}

\title{Learning to reconstruct Germano Differential Filtered Fields with
Convolutional Neural Networks for Large Eddy Simulation of Plasma Turbulence
}

\author{Jony Castagna}
\author{Josh Williams}

\affiliation{
  organization={UKRI-STFC Hartree Centre},
  addressline={Daresbury Laboratory},
  city={Warrington},
  postcode={WA4 4AD},
  country={United Kingdom}
}

\begin{abstract}
We present a novel method which combines the invertibility of the differential filter of \citet{germano_differential_1986}, with convolutional networks to accelerate plasma turbulence simulations.
Rather than learning the mapping from coarse-grid LES fields to fine-grid DNS, the proposed method learns the mapping from coarse LES fields to the differential-filtered DNS fields. 
We then use the inverse of Germano's filter to reconstruct the fine-scale details and hence calculate exactly the nonlinear terms in the governing equations and subgrid-scale terms in the LES equations.

The method is tested on the 2D Hasegawa-Wakatani (HW) and modified Hasegawa-Wakatani (mHW) plasma turbulence models at $256^2$ and $512^2$ DNS resolutions using 2, 4 and 8 times smaller LES grids in each direction. We compare the model with DNS results on energy, enstrophy, spectra, probability distributions, poloidal and radial fluxes and computational cost. Validation extends to 1000 simulation time units for HW and 100 for mHW, whose faster dynamics lead toward laminarization.

For the $256^2$ cases, the finer LES meshes preserve several integral and spectral features, whereas the coarsest mesh yields unphysical behavior. In contrast, for the $512^2$ cases, even the coarsest LES mesh preserves the turbulent character of the flow and maintains acceptable statistics for both the HW and mHW models. Moreover, we show how our methodology is superior to traditional algebraic numerical LES models, which traditionally fail to capture important design features.
Finally, benchmarks show an MPI task-time reductions of 4.82--49.45, with a maximum wall-clock speedup of $1.55\times$ when compared to full CPU socket.
\end{abstract}


\end{frontmatter}
%----------------------------------------------------------------
%
%
%
%
%
%
%
%----------------------------------------------------------------
\section{Introduction}\label{sec:Introduction}
The dynamics of plasma turbulence in the edge region of nuclear fusion reactors are best studied through Direct Numerical Simulation (DNS). However, this approach is highly prohibitive, and restricts simulations to small systems far below the resolution required for future tokamaks such as ITER, even on modern exascale hardware. Alternatively, simpler models are used in 1D or 2D with diffusion profiles taken from experiments.
One alternative for studying edge plasmas without resolving all scales of the turbulence is using Large Eddy Simulations (LES) on coarse grids, but an ad-hoc modelling of the sub-grid scales is required. In strongly magnetized plasma it is notoriously very challenging to find correct models \citep{leddy2016thesis}, as the energy injected at small length scales and propagated to large scales through an inverse cascase, as described on the work on 2D turbulence by \citet{kraichnan_inertial_1967}. This makes plasma turbulence modelling with traditional LES models unsuitable, since they tend to be mostly dissipivative towards small scales. 

One standard problem studied in plasma edge turbulence is the 2D Hasegawa-Wakatani model \citep{hasegawa_self-organization_1987}, which couples plasma-density and electrostatic-potential fluctuations to describe resistive drift-wave turbulence. To account for zonal-flow dynamics,
the modified Hasegawa-Wakatani model excludes the zonal components from the resistive coupling \citep{numata_bifurcation_2007}. This reflects the absence of a parallel-electron response for zonal modes, which are constant on a flux surface. Removing their direct resistive damping allows zonal shear flows to develop and, depending on the parameters, suppress drift-wave turbulence. 
Turbulence modelling of the HW system through eddy-viscosity models such as Smagorinsky has been studied by \citet{leddy2016thesis}. Recently, fully data-driven models of this problem have also been studied \citep{gahr2024scientific, artigues2025accelerating, greif2023physics}, but despite the encouraging results, long time integrations are still prohibitive due to the high non linearity of the problem which can lead easily to unstable results \citep{van_de_wetering_neural_2026}. Our work focuses on integrating data-driven methods within existing LES solvers to provide numerical stability.

Deconvolution operators for LES have been proposed in the past as Approximate Deconvolution Methods (ADM) by \citet{stolz_approximate_1999}. ADM reconstructs unfiltered flow variables based on the filtered variables, but this fails in capturing small turbulent details not present in the original simulation leading to a energy `pile-up' \citep{amani2025novel}.
However, recent interest in deep neural networks for computational fluid dynamics \citep{wang_recent_2024} has presented approaches for generating subgrid flow structures through (i) Generative Adversarial Networks (GANs) \citep{goodfellow_generative_2014,bode_using_2021,nista2024influence,castagna_stylegan_2024}, (ii) diffusion models \citep{ho_denoising_2020,sardar2024spectrally,shu2023physicsDM} and other methods \citep{kim_unsupervised_2021, ling2016reynolds}. Each of these methodologies has pros and cons, mainly on training robustness, accuracy and speed. In our previous work \citep{castagna_stylegan_2024}, we used StyleGAN \citep{karras_style-based_2019}, a model developed by NVIDIA for face synthesis, as a deconvolution operator for 2D turbulent flows. The algorithm used, named Style Eddy Simulation (StylES), showed a factor $10\times$ speedup vs single core DNS for $1024^2$ resolution. However, GANs such as StyleGAN often suffer from the need for a large quantity of training data to achieve valid training. Even then, reconstructing DNS fields from LES on very coarse meshes remains challenging due to the spectral bias of neural network training \citep{rahaman2019spectralbias}. This is particularly true for lossy filters such as Gaussian or top-hat kernels. 

One class of filters for LES proposed by \citet{germano_differential_1986} are known as differential filters which are invertible at a fixed resolution. This corresponds to inverting the Helmholtz equation of a filtered variable, and has been used in particle-laden flows to reconstruct the unfiltered fluid velocity in particle drag calculations \citep{park2017particle}.
However, the invertibility should be seen more as a change of variables, rather than a reduction of dimensionality. As the LES variable will be still solved on a coarse grid, the loss of information is unavoidable and the invertibility is limited \citep{sagaut_large_2006}.

In the present paper we improved our previous work on StylES \citep{castagna_stylegan_2024}, replacing the full StyleGAN architecture with a simpler U-Net based architecture, but moving the training target from the fully developed DNS fields to the Germano's filtered DNS fields (at the DNS resolution). 
The idea is that the super-resolution task is split into two parts: (i) reconstructing the missing information from the coarse-grid LES to the fine-grid LES, and (ii) reconstruction of the final DNS field from the fine-grid LES using the inverse of Germano's differential filter.
This approach allows for: a) faster super-resolution and hence time integration and b) a simplified training process.

The paper continues as follows: Sections~\ref{sec:PhysicalModel} introduce the physical model and the LES filtering approach. Section~\ref{sec:NumericalImplementation} explains the U-Net based architecture and the modifications needed to have a stable time integration. Moreover, details on the training and reconstruction methodology are provided. Section~\ref{sec:Results} presents the results and discussion on the performance. Finally, Section~\ref{sec:Conclusions} summarizes the findings and proposes future work, which will be mainly moving towards a full 3D implementation of the proposed methodology.

The code used to generate the results is available on github at \url{https://github.com/stfc/U-Net-LES} and the model weights are available on Zenodo at \url{https://doi.org/10.5281/zenodo.22973917}.
%----------------------------------------------------------------
%
%
%
%
%
%
%
%----------------------------------------------------------------
\section{Physical model}\label{sec:PhysicalModel}

\subsection{Hasegawa-Wakatani model}
In a tokamak device, one of the first simplified model which describes the transport of turbulence in the scrape-off layer is described by Hasegawa-Wakatani equations \citep{hasegawa_self-organization_1987}. This consists of a perpendicular slab to the toroidal magnetic field, with radial coordinate $x$ and poloidal coordinate $y$. The time evolution of the normalized density perturbation $n$, electrostatic potential $\phi$ and vorticity $\zeta$ are given by
\begin{align}
    \frac{\partial\zeta}{\partial t}+\{\phi,\zeta\}
    &=\alpha(\phi-n)-D_\zeta\nabla_\perp^4\zeta,
    \label{eq:hw_vorticity}\\
    \frac{\partial n}{\partial t}+\{\phi,n\}
    &=\alpha(\phi-n)-\kappa\frac{\partial\phi}{\partial y}
      -D_n\nabla_\perp^4 n.
    \label{eq:hw_density}
\end{align}
where $\{a,b\}$ is a Poisson bracket term, $\alpha$ is a constant representing the conductivity, $\kappa=-\partial_x\ln n_0$ specifies the background density-gradient drive, $D$ is a viscous dissipation coefficient for each field (we use $D_n=D_\zeta$ throughout). Fourth-order dissipation (known as hyperviscosity) removes small-scale fluctuations. 

The Poisson bracket describing the $\mathbf{E}\times\mathbf{B}$ advection is
\begin{equation}
    \{a,b\}=\partial_xa\,\partial_yb-\partial_ya\,\partial_xb,
    \label{eq:hw_poisson}
\end{equation}
and the potential is solved by
\begin{equation}
    \nabla_\perp^2\phi=\zeta.
\end{equation}

The modified HW model removes zonal components from the coupling \citep{numata_bifurcation_2007}.
By setting:
\begin{equation}
    \langle f\rangle_y=\frac{1}{L_y}\int_0^{L_y}f(x,y,t)\,dy,
    \qquad f'=f-\langle f\rangle_y.
    \label{eq:zonal_decomposition}
\end{equation}
the equations become:
\begin{align}
    \frac{\partial\zeta}{\partial t}+\{\phi,\zeta\}
    &=\alpha(\phi'-n')-D_\zeta\nabla_\perp^4\zeta,
    \label{eq:mhw_vorticity}\\
    \frac{\partial n}{\partial t}+\{\phi,n\}
    &=\alpha(\phi'-n')-\kappa\frac{\partial\phi}{\partial y}
      -D_n\nabla_\perp^4 n.
    \label{eq:mhw_density}
\end{align}
which introduces zonal modes without parallel-electron response which. Excluding their direct resistive damping permits zonal shear flows that can suppress turbulence \citep{numata_bifurcation_2007}. We use primes for non-zonal fluctuations and reserve tildes for LES-filtered fields. This distinction matters because the zonal average and the LES filter serve different purposes: the former defines the physical coupling, while the latter introduces the unresolved nonlinear terms discussed below.

\subsection{LES formulation with differential filtering}
\label{sec:GermanoDifferentialFilter}

In the LES approach, we solve for a filtered field such as
\begin{equation}
    \widetilde{f}(\boldsymbol{x},t) 
    = 
    \int_{\mathbb{R}^3} G(\boldsymbol{x}-\boldsymbol{r}; \ell) f (\boldsymbol{r}, t) d\boldsymbol{r},
\end{equation}
where $G$ is the filter kernel with characteristic width $\ell$, which satisfies $\int_{\mathbb{R}^3} G(\mathbf{r};\ell)d\mathbf{{r}}=1$.

\citet{germano_differential_1986} introduced the differential filter
\begin{equation}
    G(\boldsymbol{x}-\boldsymbol{r};\ell) 
    = 
    \frac{1}{4\pi\ell^2} 
    \frac{\exp(-|\boldsymbol{x}-\boldsymbol{r}|/\ell)}{|\boldsymbol{x}-\boldsymbol{r}|}
\end{equation}
where the singularity is removed by integration and the inverse of the integrated filter is a differential operator \citep{germano_elliptic_1986}. For the isotropic case, we define:
\begin{equation}
    \mathcal{H}_\ell=I-\ell^2\nabla_\perp^2,
    \qquad \mathcal{F}_\ell=\mathcal{H}_\ell^{-1},
    \qquad \widetilde f=\mathcal{F}_\ell f.
    \label{eq:germano_filter}
\end{equation}
Thus filtering amounts to solving a Helmholtz problem, whereas its formal inverse is locally defined as:
\begin{equation}
    (I-\ell^2\nabla_\perp^2)\widetilde f=f,
    \qquad f=\widetilde f-\ell^2\nabla_\perp^2\widetilde f.
    \label{eq:germano_inverse}
\end{equation}
The inverse requires boundary conditions consistent with the forward filter, where our implementation uses periodic Fourier transforms.

For our two-dimensional periodic domain, the filter is therefore specified by its Fourier multiplier:
\begin{equation}
    \widehat{\widetilde{f}}(\mathbf{k})
    =\frac{\widehat f(\mathbf{k})}{1+\ell^2|\mathbf{k}|^2},
    \qquad
    \widehat f(\mathbf{k})
    =(1+\ell^2|\mathbf{k}|^2)\widehat{\widetilde{f}}(\mathbf{k}).
    \label{eq:germano_transfer}
\end{equation}
where $\widehat{\cdot}$ denotes Fourier coefficients. 
The zero mode is preserved, and short wavelengths are attenuated algebraically without a sharp spectral cutoff. 
% This differs from Gaussian filtering, whose attenuation is exponential in squared wavenumber. 
% \autoref{fig:hw_filter_spectra} compares the unfiltered, Gaussian-filtered and differential-filtered energy spectra for a HW snapshot, illustrating their different attenuation at high wavenumbers. 

% Germano also gives the constant-coefficient anisotropic extension $I-A_{ij}\partial_i\partial_j$, with positive-definite $A_{ij}$ and multiplier $(1+A_{ij}k_i k_j)^{-1}$; we use the isotropic choice $A_{ij}=\ell^2\delta_{ij}$.

% \begin{figure}[htbp]
% \centering
% \includegraphics[width=0.85\linewidth]{images/fields_HW_run0_time501_N256_filter_spectra.pdf}
% \caption{Energy spectra for the HW run-0 snapshot at $t=501$ on a $256^2$ mesh, comparing the unfiltered field with Gaussian and Germano filtering. Both filters attenuate the high-wavenumber spectrum; the Gaussian-filtered spectrum falls more steeply than the Germano-filtered spectrum in the plotted comparison.}
% \label{fig:hw_filter_spectra}
% \end{figure}

For homogeneous $\ell$ on the periodic domain, filtering is linear and commutes with spatial derivatives and the time derivative. Therefore, the filtering procedure satisfies the criteria for LES filtering of linearity, conservation and commutation \citep{sagaut_large_2006}. 
It also commutes with the poloidal average defining the mHW coupling. 
These properties justify filtering the linear terms directly. 
% However, this cannot be assumed for spatially varying filter widths or incompatible boundary treatments \citep{klein2020commutation}. 

Applying the filter, $\mathcal{F}_\ell$, to \eqref{eq:hw_vorticity}-\eqref{eq:hw_density} (omitting the filtered mHW equations for brevity) yields
\begin{align}
    \partial_t\widetilde\zeta+\{\widetilde\phi,\widetilde\zeta\}
    &=\alpha(\widetilde\phi-\widetilde n)
      -\tau_{\mathrm{SGS},\zeta}-D_\zeta\nabla_\perp^4\widetilde\zeta,
    \label{eq:filtered_hw_vorticity}\\
    \partial_t\widetilde n+\{\widetilde\phi,\widetilde n\}
    &=\alpha(\widetilde\phi-\widetilde n)
      -\tau_{\mathrm{SGS},n}-\kappa\partial_y\widetilde\phi
      -D_n\nabla_\perp^4\widetilde n.
    \label{eq:filtered_hw_density}
\end{align}
where the exact nonlinear residual is
\begin{equation}
    \tau_{\mathrm{SGS},f}
    =\mathcal{F}_\ell\{\phi,f\}-\{\widetilde\phi,\widetilde f\},
    \qquad f\in\{n,\zeta\}.
    \label{eq:sgs_bracket}
\end{equation}
where the first term is calculated with our learned reconstruction. Note that in Section \ref{sec:les-models}, we benchmark this method against algebraic LES models for the turbulent (eddy) viscosity, including Smagorinsky and CENTORI \citep{smagorinsky1963general, knight2012centori, leddy2016thesis}.

A central result of \citet{germano_differential_1986} is that the differential inverse gives exact expressions for filtered correlations. For two smooth scalar fields $a$ and $b$, define $C_\ell(a,b)=\mathcal{F}_\ell(ab)-\widetilde a\widetilde b$. The product rule and Eq.~\eqref{eq:germano_inverse} gives:
\begin{equation}
    \mathcal{H}_\ell C_\ell(a,b)
    =2\ell^2\nabla_\perp\widetilde a\cdot\nabla_\perp\widetilde b
     +\ell^4(\nabla_\perp^2\widetilde a)(\nabla_\perp^2\widetilde b).
    \label{eq:germano_correlation}
\end{equation}
Consequently, the exact correlation is the forward filter of the right-hand side, rather than just its leading gradient term. Applied to the two products in the Poisson bracket, this identity gives
\begin{equation}
    \tau_{\mathrm{SGS},f}
    =C_\ell(\partial_x\phi,\partial_y f)
     -C_\ell(\partial_y\phi,\partial_x f).
\end{equation}
Equivalently, the nonlinear term can be evaluated by reconstructing both fields:
\begin{equation}
    \mathcal{F}_\ell\{a,b\}
    =\mathcal{F}_\ell\{
       \mathcal{H}_\ell\widetilde a,
       \mathcal{H}_\ell\widetilde b\}.
    \label{eq:germano_bracket_inverse}
\end{equation}
where $a=\phi$ and $b$ is $n$ or $\zeta$.
These are continuous identities, not a closure assumption or a guarantee of exact reconstruction from coarse samples. As Germano emphasizes, an exactly inverted filtered system is formally a change of dependent variables; numerical resolution still limits the information retained. 
In particular, the physical filter and the downsampling to the LES mesh are separate operations. 

Writing $\mathcal{R}_h$ for Fourier downsampling to that mesh, the network input is $(n,\phi)_{\mathrm{LES}}=\mathcal{R}_h\mathcal{F}_\ell(n,\phi)_{\mathrm{DNS}}$. Downsampling discards modes that the differential inverse alone cannot restore. Our learned reconstruction supplies an estimate of the filtered field on the fine mesh before inversion:
\begin{equation}
    (\widetilde{n},\widetilde{\phi})_{\mathrm{LES}}
    \xrightarrow{\mathcal{N}_\theta}
    (\widetilde n^{\mathrm{pred}},\widetilde\phi^{\mathrm{pred}})_{\mathrm{fDNS}}
    \xrightarrow{\mathcal{H}_\ell}
    (n^{\mathrm{rec}},\phi^{\mathrm{rec}})_{\mathrm{DNS}}.
    \label{eq:germano_learning_path}
\end{equation}
The inverse multiplier grows as $|\mathbf{k}|^2$, amplifying high-wavenumber prediction errors. The implementation therefore permits spectral regularization before or after inversion; when enabled, it changes the reconstruction from the exact inverse in Eq.~\eqref{eq:germano_inverse}. Vorticity is derived consistently as $\zeta^{\mathrm{rec}}=\nabla_\perp^2\phi^{\mathrm{rec}}$. The reconstructed nonlinear terms are then filtered and restricted,
\begin{equation}
    \mathcal{F}_\ell\{\phi,f\}
    =\mathcal{R}_h\mathcal{F}_\ell
      \{\phi^{\mathrm{rec}},f^{\mathrm{rec}}\},
      \qquad f\in\{n,\zeta\}.
\end{equation}
providing the LES solver with an approximation to the filtered fine-grid advection. This separates the analytically defined Germano inverse from the learned recovery of information lost through mesh downsampling.
%----------------------------------------------------------------
%
%
%
%
%
%
%
%----------------------------------------------------------------
\section{Numerical Implementation}\label{sec:NumericalImplementation}
We describe here the numerical implementation details for the full methodology. The settings below refer to the HW reconstruction from $32^2$ LES $\to$ $256^2$ DNS; they are not assumed to apply to every resolution or modified-HW case, but give very good guidance for other configurations. We used BOUT++ solver \citep{dudson_bout++_2009} for solving the filtered HW and mHW equations. In the two-dimensional runs, the paper's poloidal coordinate $y$ corresponds to the code's $z$ coordinate. For further details, see the Hasegawa-Wakatani example in the BOUT++ repository.

\subsection{NN-Architecture}\label{sec:NN-Architecture}
We use a modification of the variational U-Net architecture \citep{ronneberger_u-net_2015, esser2018variational} for the reconstruction model, hereafter referenced as \textbf{DirectUNetSR}. While in U-Net encoder and decoder are very similar in structure, the DirectUNetSR is more asymmetric. Being the input at LES coarse resolution, we only use a small encoder followed by a decoder to upsample the input to the fine DNS resolution. Skip connections are only used at low levels as we found that they tend to introduce artifacts during time integration which eventually degrade the reconstruction quality and lead to very strong oscillations. Despite skip connections prevent generative behaviour, they are essential for a good reconstruction as they carry important information of the input LES configuration.

\begin{figure}[htbp]
\centering
\includegraphics[width=\linewidth]{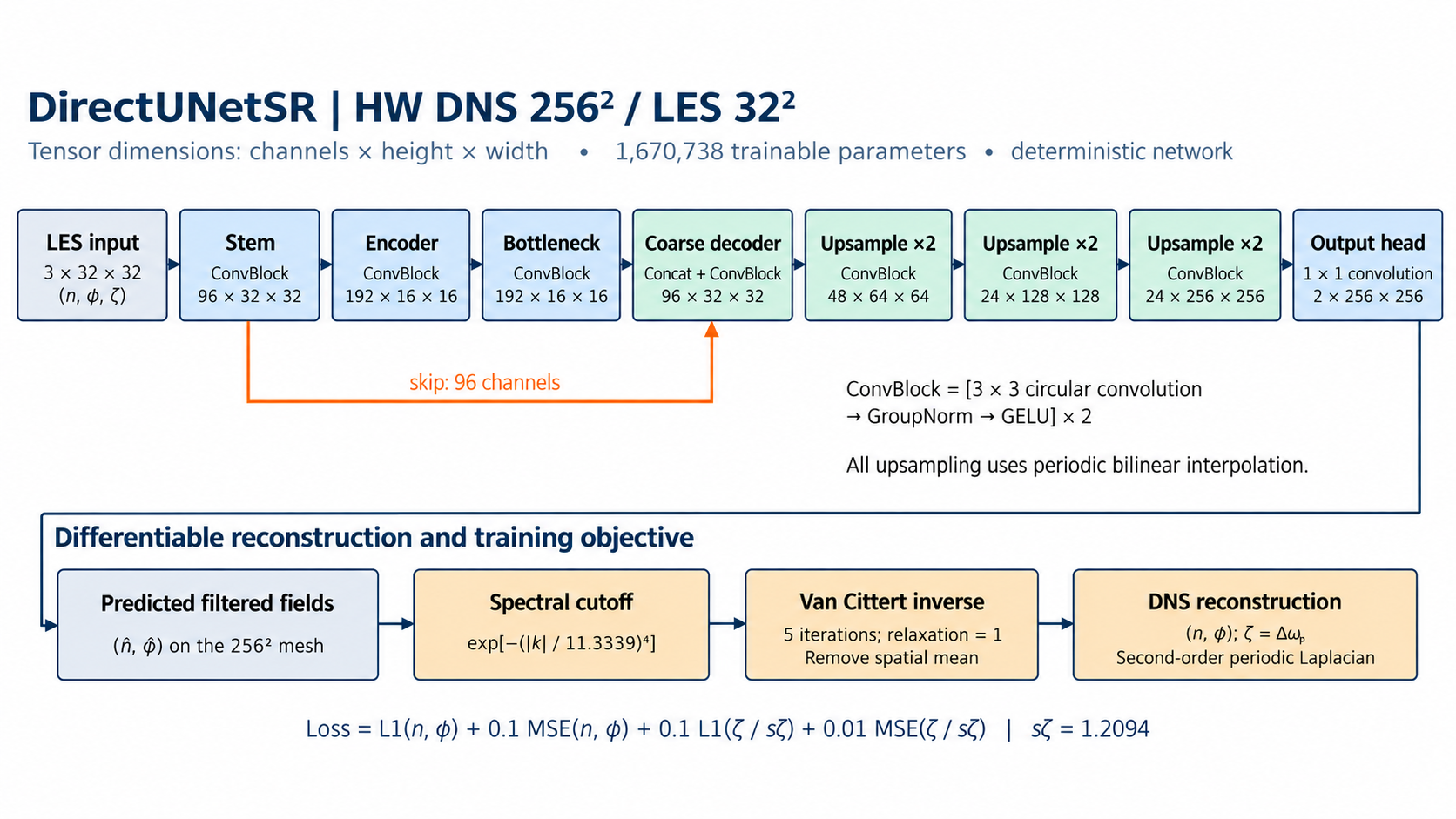}
\caption{Current \texttt{DirectUNetSR} architecture for the HW $32^2$ LES $\to$ $256^2$ DNS training configuration. Tensor sizes are channels $\times$ height $\times$ width, omitting the batch dimension. The orange connection concatenates the stem features with the upsampled bottleneck features. Blue and green blocks contain learned convolutions; the spectral cutoff, van Cittert iterations and vorticity evaluation are fixed differentiable operations.}
\label{fig:direct_unet_architecture}
\end{figure}

The three coarse input channels are density, potential and vorticity, denoted by $\mathbf{U}=(n,\phi,\zeta)$. The network predicts two fine-grid filtered fields, $\widetilde n$ and $\widetilde\phi$, being the vorticity obtained from the Laplacian of the potential fields $\zeta=\nabla_\perp^2\phi$. The encoder begins with 96 feature channels on the $32^2$ input mesh and uses a 192-channel bottleneck on a $16^2$ mesh. A skip-connected decoder returns to $32^2$, followed by periodic bilinear upsampling stages reaching $64^2$, $128^2$ and $256^2$. Periodic padding was applied prior to convolution and interpolation operations to the domain topology. The model has 1,670,738 trainable parameters. 
Figure~\ref{fig:direct_unet_architecture} summarizes the network and its differentiable reconstruction path.

\subsection{Reconstruction}
The output from the DirectUNetSR model is passed through a smooth spectral cutoff with multiplier to eliminate excessively high wavenumbers. The cutoff is configured to remove modes above the Nyquist frequency of the coarse LES mesh, with a smooth transition to avoid Gibbs oscillations. The cutoff multiplier is
\begin{equation}
    Q(\mathbf{k})=\exp\left[-\left(\frac{|\mathbf{k}|}{k_c}\right)^8\right],
    \qquad k_c=0.707\,\frac{\pi}{h}\simeq11.3339.
\end{equation}
The value of $k_c=0.707$ is chosen to match the Nyquist frequency of the DNS energy spectra (considered null at $10^{-10}$). For stability purposes, the configured inverse is a truncated van Cittert reconstruction \citep{vancittert1931}, rather than directly apply the inverse. For $g=Q\widetilde f_\theta$, implemented spectrally, it computes
\begin{equation}
    f^{(0)}=g,\qquad
    f^{(j+1)}=f^{(j)}+\omega\left[g-\mathcal{F}_\ell f^{(j)}\right],
    \quad j=0,\ldots,4,\qquad\omega=1.
    \label{eq:training_van_cittert}
\end{equation}
where $\omega$ is a relaxation parameter. The iteration is differentiable, and the network can learn to compensate for the approximate inverse. The reconstructed field is $f^{(5)}$ with its spatial mean removed to satisfy the divergence-free constraint on mass and electric charge. Thus this run uses five iterations of a regularized approximate inverse, rather than the exact differential inverse in Eq.~\eqref{eq:germano_inverse}. No post-inversion filter is enabled. Reconstructed vorticity is calculated from the potential using the second-order periodic finite-difference Laplacian,
\begin{equation}
    \zeta_\theta=\Delta_h\phi_\theta,\qquad
    (\Delta_h\phi)_{i,j}
    =\frac{\phi_{i+1,j}+\phi_{i-1,j}+\phi_{i,j+1}+\phi_{i,j-1}-4\phi_{i,j}}{h^2},
\end{equation}
with periodic indices. The DNS vorticity target is likewise derived from its potential. We use this hard constraint to strongly enforce physical consistency of the learned vorticity and potential fields.

\begin{figure}[htbp]
\centering
\includegraphics[width=0.85\linewidth]{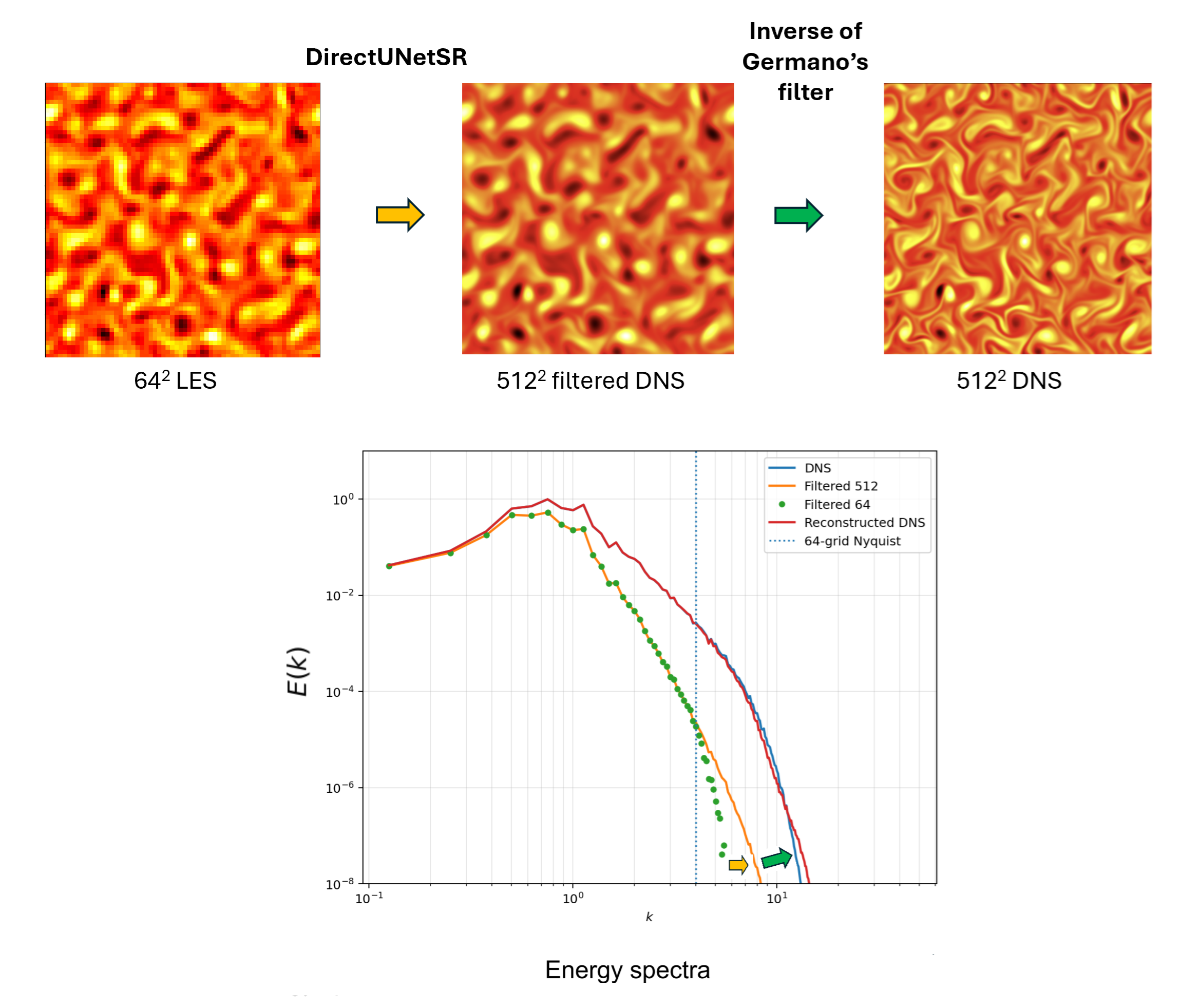}
\caption{Reconstruction and inversion of the vorticity field for the HW with DNS resolution $512^2$. The energy spectra shows the DNS value, the filtered fields on the $512^2$ and $64^2$ grids, and the reconstructed DNS. The dotted vertical line marks the Nyquist wavenumber of the $64^2$ grid.}
\label{fig:reconstruction_energy_spectrum}
\end{figure}

Figure~\ref{fig:reconstruction_energy_spectrum} shows the methodology applied to the vorticity field for the HW test case at $512^2$ resolution from $64^2$ LES input resolution, comparing the DNS, LES input, filtered and reconstructed energy spectra.

\subsection{Loss function}
The objective function compares the reconstructed fields with DNS after cutoff, inversion and vorticity evaluation; gradients propagate through this entire path and it does not directly penalize the network's intermediate filtered output. Let:
\begin{equation}
    \overline{n}=\frac{n_\theta-n_{\mathrm{DNS}}}{\sigma_n},\qquad
    \overline{\phi}=\frac{\phi_\theta-\phi_{\mathrm{DNS}}}{\sigma_\phi},\qquad
    \overline{\zeta}=\frac{\zeta_\theta-\zeta_{\mathrm{DNS}}}{\sigma_\zeta},
\end{equation}
be the normalized reconstruction errors, where $f_\theta$ denotes the reconstructed field and $f_{\mathrm{DNS}}$ its DNS target for each quantity $f\in\{n,\phi,\zeta\}$. $\sigma_f$ is the loss normalization scale, given by the training-data standard deviation when that normalization is enabled and otherwise by a unit scale. With $\mu_{B,\Omega}[q]$ denoting the statistical mean of $q$ over the batch and spatial grid, define
\begin{align}
    \mathcal{L}_{n\phi}^{(1)}
    &=\tfrac12\mu_{B,\Omega}\!\left[|\overline{n}|+|\overline{\phi}|\right],&
    \mathcal{L}_{n\phi}^{(2)}
    &=\tfrac12\mu_{B,\Omega}\!\left[\overline{n}^2+\overline{\phi}^2\right],\\
    \mathcal{L}_{\zeta}^{(1)}
    &=\mu_{B,\Omega}\!\left[|\overline{\zeta}|\right],&
    \mathcal{L}_{\zeta}^{(2)}
    &=\mu_{B,\Omega}\!\left[\overline{\zeta}^2\right].
\end{align}
the implemented loss is defined as:
\begin{align}
    \mathcal{L}
    &=\lambda_{n\phi}^{(1)}\mathcal{L}_{n\phi}^{(1)}
      +\lambda_{n\phi}^{(2)}\mathcal{L}_{n\phi}^{(2)}
      +\lambda_{\zeta}^{(1)}\mathcal{L}_{\zeta}^{(1)}
      +\lambda_{\zeta}^{(2)}\mathcal{L}_{\zeta}^{(2)}.
    \label{eq:training_loss}
\end{align}
where $\lambda_{n\phi}^{(1)}$, $\lambda_{n\phi}^{(2)}$, $\lambda_\zeta^{(1)}$ and $\lambda_\zeta^{(2)}$ denote the respective loss weights.

The density/potential terms average the two channels, so each coefficient weights their combined mean loss. The squared-error contribution increases the penalty on larger residuals, while the vorticity terms constrain potential curvature through the derived field.

\begin{table}[ht]
\centering
\caption{Loss settings recorded for the HW $256^2\rightarrow32^2$ training configuration. The four effective coefficients in Eq.~\eqref{eq:training_loss} correspond to the density/potential L1 and MSE weights and the vorticity L1 and MSE weights, respectively; the implementation forms the last coefficient as the product of its recorded vorticity and MSE weights.}
\label{tab:training_loss_weights}
\begin{tabular}{lll}
\toprule
Coefficient & Value & Role \\
\midrule
$\lambda_{n\phi}^{(1)}$ & 1.0 & Unit density/potential L1 prefactor \\
$\lambda_{n\phi}^{(2)}$ & 0.1 & Density/potential MSE weight \\
$\lambda_\zeta^{(1)}$ & 0.1 & Normalized vorticity L1 weight \\
$\lambda_\zeta^{(2)}$ & 0.01 & Normalized vorticity MSE weight \\
\bottomrule
\end{tabular}
\end{table}

\subsection{Dataset}\label{sec:Dataset}

The DNS dataset were generated using the BOUT++ plasma solver. 
Table \ref{tab:cases}
and \ref{tab:dataset_effort} summarises the test case resolutions, physical parameters
and computational efforts from \citet{castagna_stylegan_2024}. Four configurations were analysed, two for the HW model and two for the mHW model, with $256^2$ and $512^2$ DNS meshes. 
Both HW and mHW systems are considered fully periodic in $x$ and $y$, with domain size $L_x=L_y=50.176$. The background density gradient is adjusted according to the test case, $\kappa=0.5 - 1$, while the dissipation coefficients are set to $D_\zeta=D_n=10^{-3}$. The parallel-electron response is $\alpha=0.5 - 1$, which are ideal values for the weakly collisional regime. 
The initial conditions are a combination (mixture) of Fourier modes. 

The different dataset have a different final time $T_f$ but the same number of trajectories $N_{runs}$. We start collecting samples from $T_s$ every $\Delta T_s$ for a total $N_s$ samples. The choice of $\Delta T_s$ is such that the variation in Similarity Structural Index Metric (SSIM) is always $\geq$ 10\%. The last 2 columns of Table \ref{tab:dataset_effort} show the overall computational cost $C_{cost}$ in node per hours, where 1 node of ScafellPike supercomputer at Hartree Centre contains a 2 socket of 16-cores Intel Xeon Gold processors, and the disk space occupied by the data.

\begin{table}[htbp]
\centering
\caption{Physical parameters for each test case.}
\label{tab:cases}
\begin{tabular}{l|c c c c}
\hline
Type & $N$ & $\alpha$ & $\kappa$ & $\nu_\zeta=\nu_n$ \\
\hline
HW  & $256^2$ & 1 & 0.5 & $10^{-3}$ \\
HW  & $512^2$ & 1 & 1   & $10^{-3}$ \\
mHW & $256^2$ & 1 & 1   & $10^{-3}$ \\
mHW & $512^2$ & 1 & 1   & $10^{-3}$ \\
\hline
\end{tabular}
\end{table}

\begin{table}[htbp]
\centering
\caption{Computational cost and memory usage for each test case.}
\label{tab:dataset_effort}
\begin{tabular}{l|c c c c c c c c }
\hline
Type & $N$ & $T_s$ & $T_f$ & $\Delta T_s$ & $N_{\mathrm{runs}}$ & $N_s$ & $C_{\mathrm{cost}}$ & disk [GB] \\
\hline
HW  & $256^2$ & 501 & 1000 & 10 & 100 & 5000 & 5.8 & 7.4 \\
HW  & $512^2$ & 201 & 1000 & 10 & 100 & 8000 & 111 & 47 \\
mHW & $256^2$ & 201 & 400  & 4  & 100 & 5000 & 5.8 & 7.4 \\
mHW & $512^2$ & 201 & 400  & 4  & 100 & 5000 & 111 & 30 \\
\hline
\end{tabular}
\end{table}

\subsection{Training}\label{sec:Training}
We split the dataset into 80\% for training, 10\% for validation and the rest 10\% for testing.
The run for 100 epochs with batch size 4, using AdamW at a constant learning rate of $10^{-4}$, weight decay $10^{-6}$, moment coefficients $(0.9,0.999)$ and $\epsilon=10^{-8}$. Gradient norms are clipped to 1.0. Computation uses float64 precision on an NVIDIA H100 80GB GPU. Synchronized random periodic shifts augment the training fields while preserving alignment of inputs and targets.

\begin{figure}[htbp]
\centering
% Show only the Total Training Objective panel of the original 180-dpi summary.
\includegraphics[width=0.75\linewidth,trim=736bp 688bp 372bp 42bp,clip]{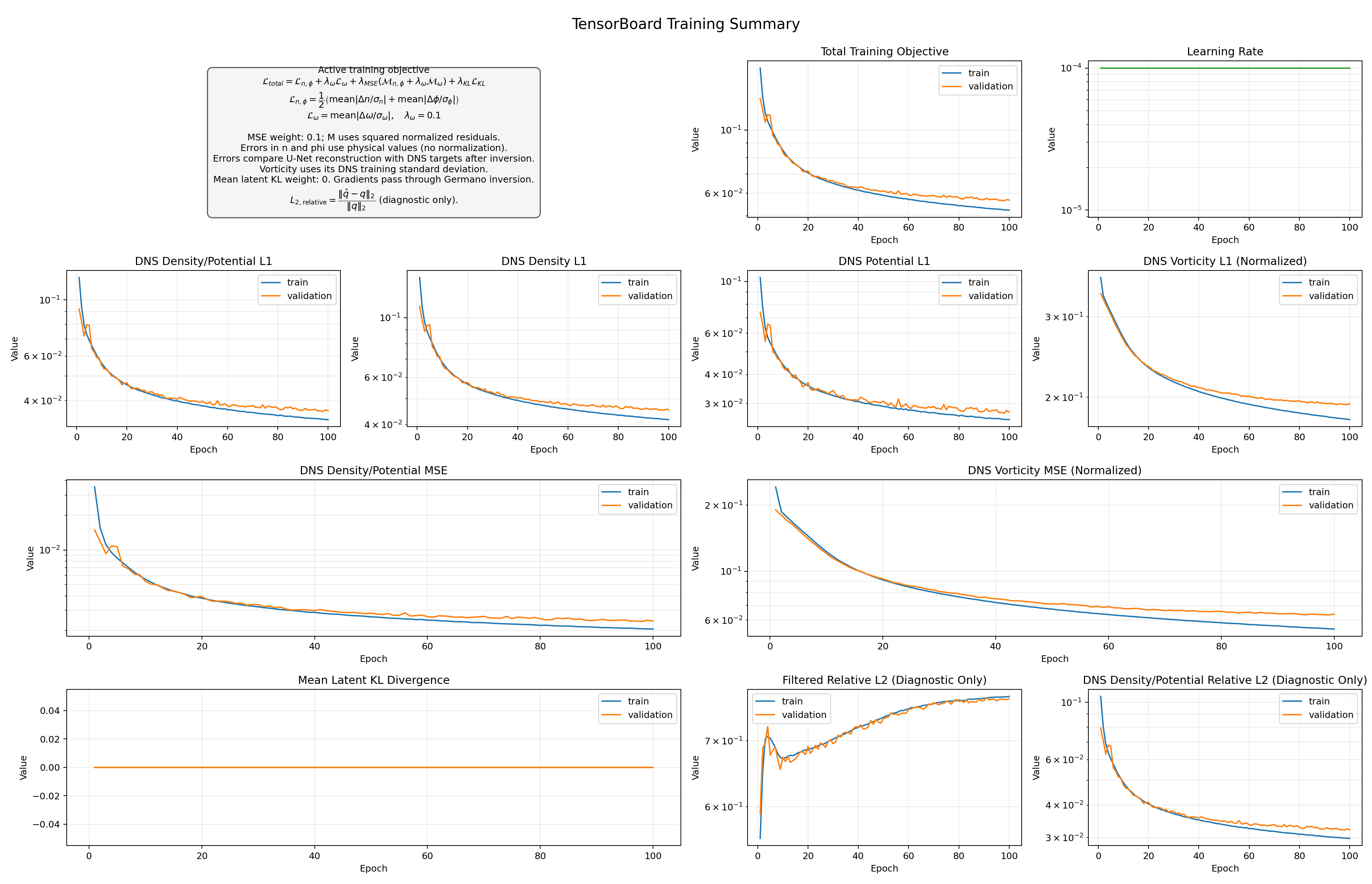}
\caption{Total training and validation objective, Eq.~\eqref{eq:training_loss}, for the HW configuration with $256^2$ DNS and $32^2$ LES grids over 100 epochs.}
\label{fig:training_objective}
\end{figure}

Figure~\ref{fig:training_objective} shows the loss as function of number of epochs for the 
$32^2$ LES $\to$ $256^2$ DNS configuration. A rapid decrease in the total objective during the first 20 epochs, followed by a slower reduction throughout the remaining training. At epoch 100, the training and validation objectives are approximately $5.2\times10^{-2}$ and $5.6\times10^{-2}$, respectively. The validation curve closely follows the training curve initially and develops a small gap at later epochs, while continuing to decrease overall. This behaviour indicates diminishing gains from further optimization without a sustained increase in validation loss over the recorded interval.

%----------------------------------------------------------------
%
%
%
%
%
%
%
%----------------------------------------------------------------
\section{Results}\label{sec:Results}

We compare HW and mHW at DNS resolutions of $256^2$ and $512^2$, each with three LES resolutions. The $256^2$ cases use $128^2$, $64^2$ and $32^2$ LES meshes; the $512^2$ cases use $256^2$, $128^2$ and $64^2$ meshes. These correspond to reductions by factors of two, four and eight in each direction, respectively. The simulations are run up to 1000 time units ($t_{HW}=\Omega_{ci}t_{physical}$, where $\Omega_{ci}=\frac{ZeB}{m_i}$ is the ion cyclotron frequency) for all HW cases and 100 time units for mHW due to the faster dynamics leading to laminarisation. 
Snapshots are saved every 1 time unit.

The statistical figures show energy, enstrophy and particle-flux histories, final-time spectra and three final-time PDFs. The mHW\_N256 plots are regenerated from the DNS dumps and saved U-Net reconstructed fields for the $128^2$, $64^2$ and $32^2$ LES meshes, using the same $0\leq t\leq100$ window. The comparisons are against DNS; no separate conventional-closure curve is present. However, before diving into into our approach results, we will show why traditional LES models fail to capture the correct nature of the turbulence.

\subsection{Comparison with algebraic LES models}\label{sec:les-models}
We compare three algebraic LES models for the HW system on a $64^2$ mesh against the $256^2$ DNS: Smagorinsky, CENTORI and implicit LES (no subgrid model). All use $\alpha=1$, $\kappa=0.5$, a periodic domain of size $50.176\times50.176$ and are initialized from the same filtered DNS snapshot. No neural reconstruction is used in these runs. The Smagorinsky model applies strain-dependent diffusion to density and vorticity, with model coefficient $C_s=0.3$. CENTORI instead uses a local diffusivity $\nu_C=C_C\{[\alpha(\phi-n)]^2+\zeta^2\}$, with $C_C=0.08$, multiplying the Laplacian of each field (see \cite{leddy2016thesis} for more details). Both explicit closures retain background fourth-order diffusion with $D_n=D_\zeta=10^{-3}$. The implicit configuration omits both the explicit closure and background diffusion, leaving its small-scale behaviour controlled by the numerical discretization.

\begin{figure}[htbp]
\centering
\begin{subfigure}[t]{0.48\textwidth}
\includegraphics{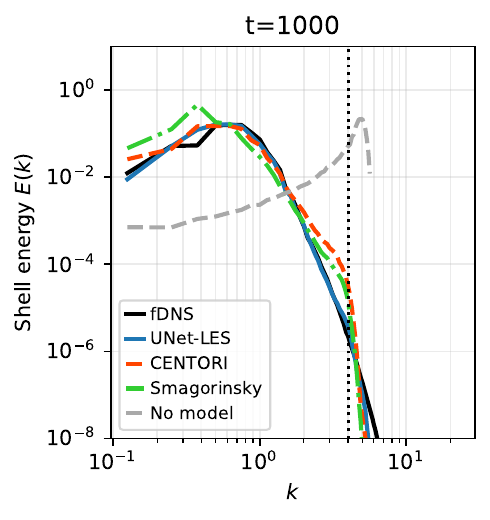}
\end{subfigure}
\begin{subfigure}[t]{0.48\textwidth}
\includegraphics{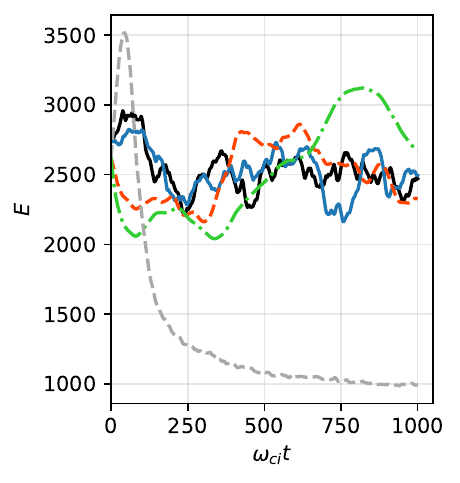}
\end{subfigure}
\caption{
Energy statistics at LES level for all reference models. Panels show the final shell energy spectrum (left) and domain-averaged energy with time (right) for $256^2$ filtered DNS (fDNS) to $64^2$ LES resolution. The dotted vertical line marks the $64^2$-grid Nyquist wavenumber.
}
\label{fig:les-models-spectra}
\end{figure}

Figure~\ref{fig:les-models-spectra} shows the energy time-evolution and the final shell energy spectra. Our DirectUNetSR model (at LES resolution) matches the filtered DNS spectrum perfectly, demonstrating excellent stability for long time integrations. Among the algebraic subgrid models, CENTORI matches the filtered DNS well at large scales until around $k=2$ where energy begins to accumulate towards the Nyquist cutoff wavenumber. The Smagorinsky model mistmatches the energy mostly, and implicit LES shows significant disagreement at low wavenumbers and a pronounced accumulation near the grid scale, with a peak just beyond the marked $64^2$-grid Nyquist wavenumber. It does not preserve the DNS spectral distribution over this interval. Observing the time evolution (Figure~\ref{fig:les-models-spectra} right), the implicit LES decays quickly to zero energy, showing the flow has laminarised due to lack of correct dissipation to sustain the turbulence. Smagorinsky and CENTORI models sustain a similar level of energy, but show less frequency in fluctuations compared to the filtered DNS. The DirectUNetSR matches fluctuating behaviour of the DNS well throughout the simulation.

\begin{figure}[htbp]
\centering
\includegraphics{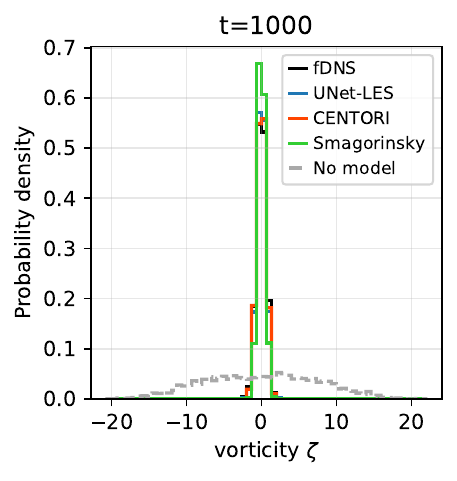}
\caption{Initial ($t=0$, left) and final ($t=1000$, right) spatial vorticity PDFs for the $256^2$ DNS and three $64^2$ LES configurations.}
\label{fig:les-models-vorticity}
\end{figure}

The vorticity PDFs in Figure~\ref{fig:les-models-vorticity} provide a complementary comparison. Initially, the LES distributions nearly coincide and have a slightly higher central peak than DNS. At $t=1000$, Smagorinsky and CENTORI remain close to the DNS distribution, reproducing its narrow central region with modest differences in the peak and shoulders. Implicit LES instead develops a much broader distribution and a markedly lower central probability density, consistent with the excessive small-scale content visible in its spectrum. 

\begin{figure}[htbp]
\centering
\includegraphics{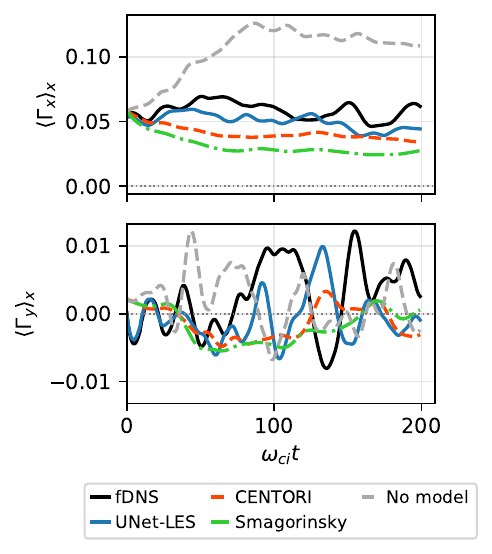}
\caption{Domain-averaged radial (upper panel) and poloidal (lower panel) turbulent particle fluxes for the $256^2$ HW filtered DNS, our UNet model, Smagorinsky, CENTORI and implicit LES on a $64^2$ mesh over $0\leq t\leq1000$. Density fluctuations are defined relative to the poloidal mean.}
\label{fig:les-models-particle-flux}
\end{figure}

Figure~\ref{fig:les-models-particle-flux} shows the particle flux statistics. Such transport statistics are the most important quantity for engineering design. To improve visualisation of the overlapping and chaotic timeseries, we (i) zoomed into the first 200 time units for HW cases, and (ii) follow the procedure of \citet{leddy2016thesis} by applying a Savitzky-Golay filter \citep{savitzky1964smoothing} with a window of 20 steps and polynomial order 3 for smoothing for all cases. 

The fluxes are clearly mispredicted by LES with no model. The CENTORI and Smagorinsky models again do not reproduce the correct fluctuating behaviour found in the filtered DNS and our DirectUNetSR model. The Smagorinsky radial flux is dissipated to around 40\% of the DNS value. Similarly, the poloidal fluxes differ from DNS in the timing, amplitude and sign of individual events. The implicit solution progressively suppresses these fluctuations and stays close to zero at late times. These transport diagnostics complement the spectra and vorticity PDFs: explicit closures preserve sustained transport more effectively than implicit LES in this case, but agreement in the spectral peak or central PDF does not ensure accurate radial-flux levels or instantaneous poloidal fluxes. The reference algebraic models are not able to reproduce the correct dynamics, compared to the DirectUNetSR model which shows excellent agreement with the filtered DNS.

\subsection{Spatial reconstruction}

\begin{figure}[p]
\centering
\includegraphics[width=\linewidth]{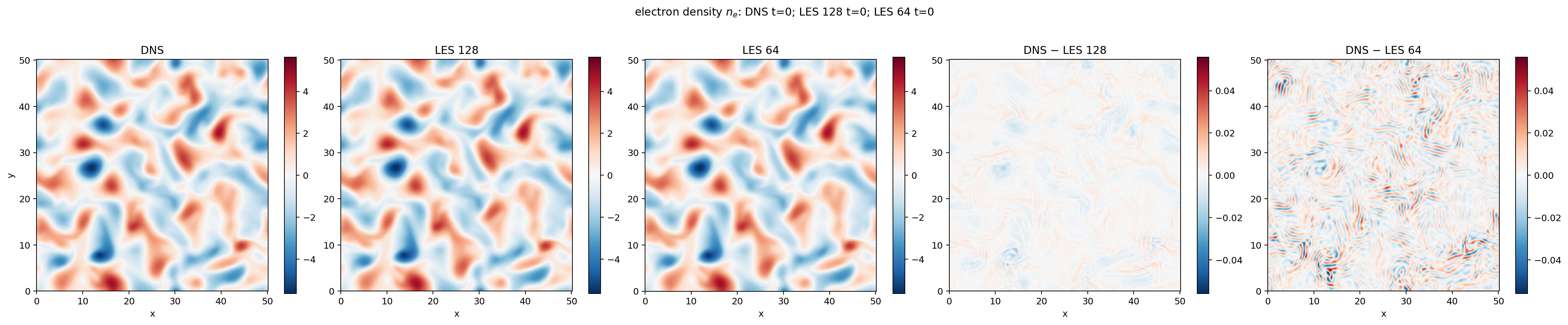}
\par\smallskip
\includegraphics[width=\linewidth]{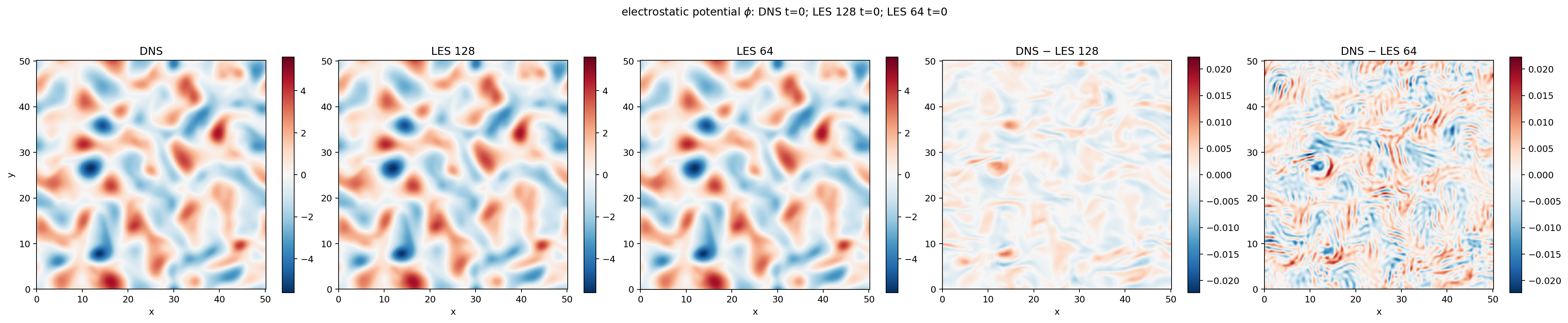}
\par\smallskip
\includegraphics[width=\linewidth]{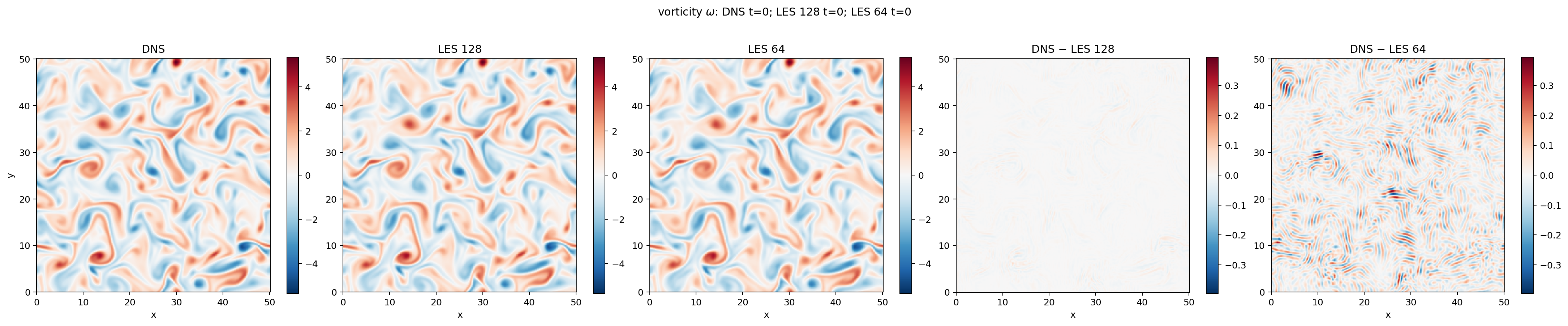}
\par\smallskip
\caption{HW\_N256 spatial comparisons at $t=0$: density (top), potential (middle) and vorticity (bottom). Each row shows DNS, the two finer LES reconstructions and their signed differences from DNS, as labelled. Signed differences are defined as DNS minus LES.}
\label{fig:fields-HW_N256}
\end{figure}

\begin{figure}[p]
\centering
\includegraphics[width=\linewidth]{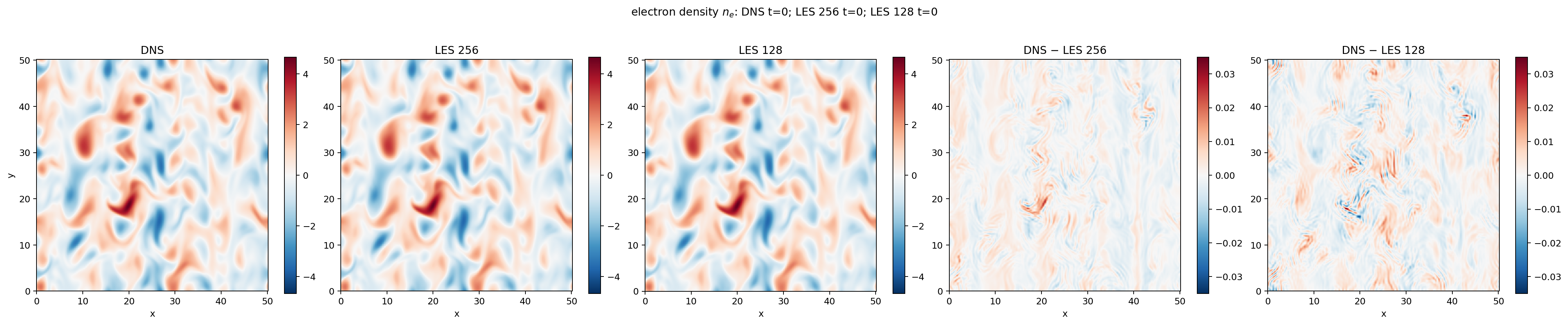}
\par\smallskip
\includegraphics[width=\linewidth]{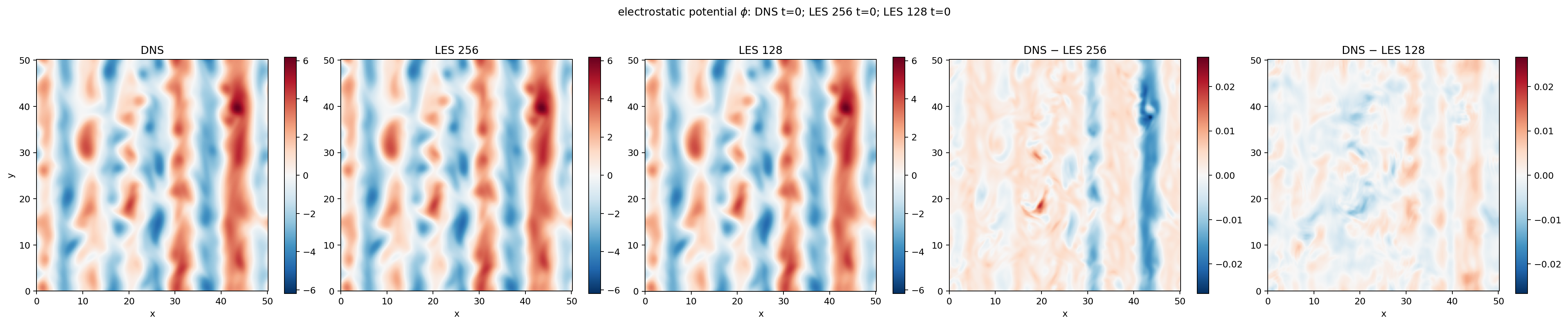}
\par\smallskip
\includegraphics[width=\linewidth]{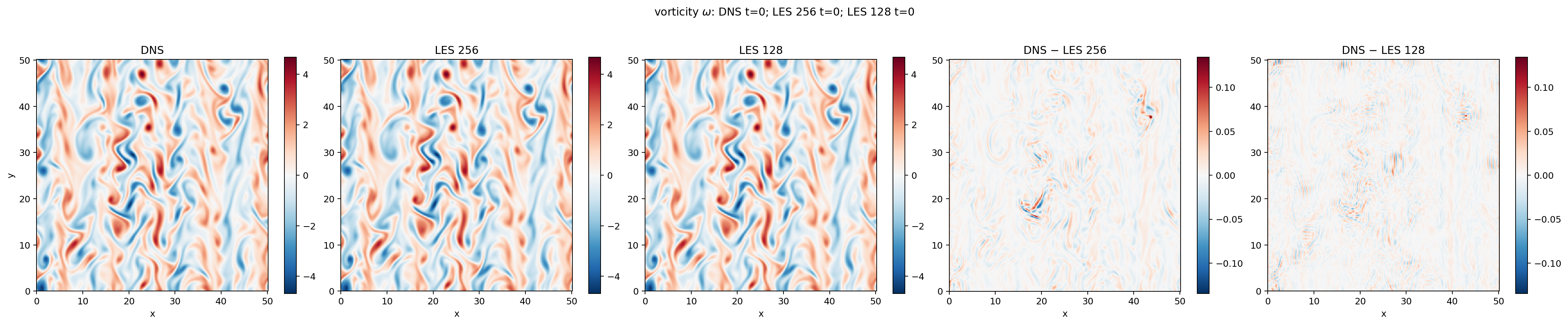}
\par\smallskip
\caption{mHW\_N512 spatial comparisons at $t=0$: density (top), potential (middle) and vorticity (bottom). Each row shows DNS, the two finer LES reconstructions and their signed differences from DNS, as labelled. Signed differences are defined as DNS minus LES.}
\label{fig:fields-mHW_N512}
\end{figure}

Figures~\ref{fig:fields-HW_N256}--\ref{fig:fields-mHW_N512} compare the initial density, potential and vorticity fields at $t=0$ for DNS and the two finer LES meshes. These comparisons assess the spatial fidelity of the initial reconstruction. The HW panels resolve vortex structures, while the mHW potential panels show bands extended in the poloidal direction. The signed difference panels, defined as DNS minus LES, identify local reconstruction errors in each field. In all cases, the highest pointwise error is in the vorticity field due to (1) the finer spatial structures being more difficult to capture, and (2) the field is not explicitly learned, it is instead reconstructed from the consistency relation $\zeta=\nabla^2\phi$.

% Only the initial spatial comparisons are shown here. They provide a qualitative assessment of reconstruction fidelity but do not establish dynamical accuracy or an integrated error norm. The subsequent evolution is assessed through the statistical diagnostics below. The coarsest LES meshes are included in those diagnostics, but are absent from the archived spatial comparison panels.

% \begin{figure}[p]
% \centering
% \includegraphics[width=\linewidth]{images/validation_current/HW_N512/U_comparison_frames/U_0000000.png}
% \par\smallskip
% \includegraphics[width=\linewidth]{images/validation_current/HW_N512/V_comparison_frames/V_0000000.png}
% \par\smallskip
% \includegraphics[width=\linewidth]{images/validation_current/HW_N512/P_comparison_frames/P_0000000.png}
% \par\smallskip
% \caption{HW\_N512 spatial comparisons at $t=0$: density (top), potential (middle) and vorticity (bottom). Each row shows DNS, the two finer LES reconstructions and their signed differences from DNS, as labelled. Signed differences are defined as DNS minus LES.}
% \label{fig:fields-HW_N512}
% \end{figure}

% \begin{figure}[p]
% \centering
% \includegraphics[width=\linewidth]{images/validation_current/mHW_N256/U_comparison_frames/U_0000000.png}
% \par\smallskip
% \includegraphics[width=\linewidth]{images/validation_current/mHW_N256/V_comparison_frames/V_0000000.png}
% \par\smallskip
% \includegraphics[width=\linewidth]{images/validation_current/mHW_N256/P_comparison_frames/P_0000000.png}
% \par\smallskip
% \caption{mHW\_N256 spatial comparisons at $t=0$: density (top), potential (middle) and vorticity (bottom). Each row shows DNS, the two finer LES reconstructions and their signed differences from DNS, as labelled. Signed differences are defined as DNS minus LES.}
% \label{fig:fields-mHW_N256}
% \end{figure}

\subsection{Energy and enstrophy}

The diagnostics are the spatially integrated HW energy ($E$) and generalized enstrophy ($Z$),
\begin{align}
 E &= \frac12\int_\Omega\left(n^2+\nabla_x\phi^2 + \nabla_y\phi^2\right)\,d\mathbf{\Omega},\\
 Z &= \frac12\int_\Omega(n-\zeta)^2\,d\mathbf{\Omega}.
\end{align}
where $\nabla \phi$ is computed with fourth-order finite-difference derivatives. 

\begin{figure}[p]
\centering
\begin{subfigure}[t]{0.48\textwidth}
\centering
\includegraphics[width=\linewidth]{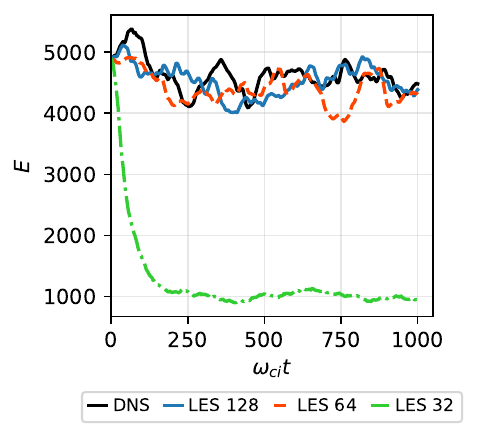}
\caption{HW\_N256}
\end{subfigure}
\hfill
\begin{subfigure}[t]{0.48\textwidth}
\centering
\includegraphics[width=\linewidth]{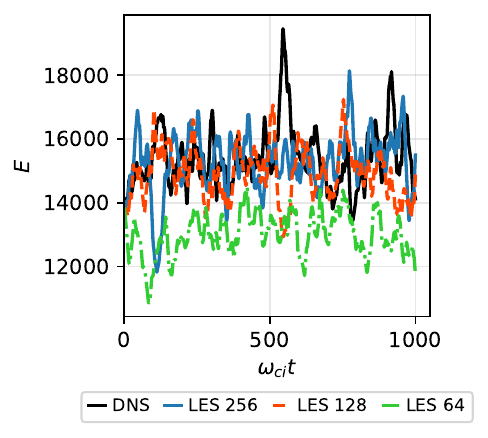}
\caption{HW\_N512}
\end{subfigure}
\par\smallskip
\begin{subfigure}[t]{0.48\textwidth}
\centering
\includegraphics[width=\linewidth]{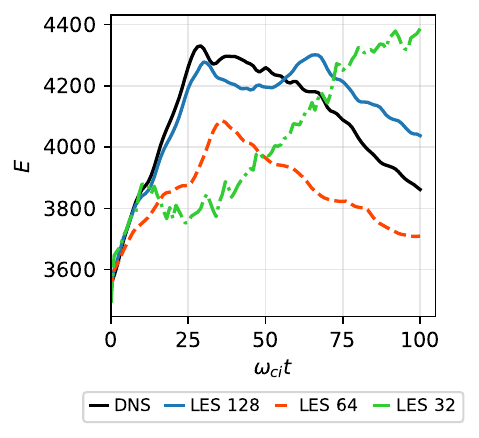}
\caption{mHW\_N256}
\end{subfigure}
\hfill
\begin{subfigure}[t]{0.48\textwidth}
\centering
\includegraphics[width=\linewidth]{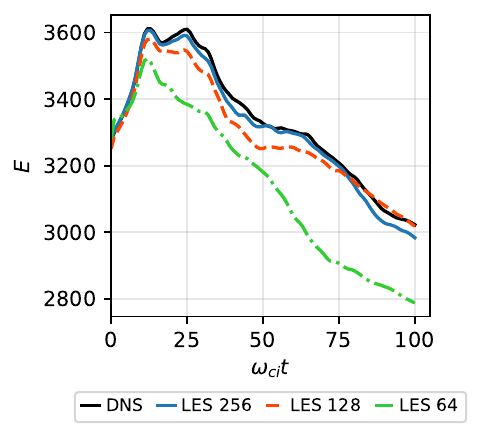}
\caption{mHW\_N512}
\end{subfigure}
\caption{Spatially integrated HW energy over $0\leq t\leq1000$ for HW and $0\leq t\leq100$ for mHW. Each available panel includes DNS and all three LES resolutions; energy uses cell-area quadrature on the diagnostic grid.}
\label{fig:energy}
\end{figure}

\begin{figure}[p]
\centering
\begin{subfigure}[t]{0.48\textwidth}
\centering
\includegraphics[width=\linewidth]{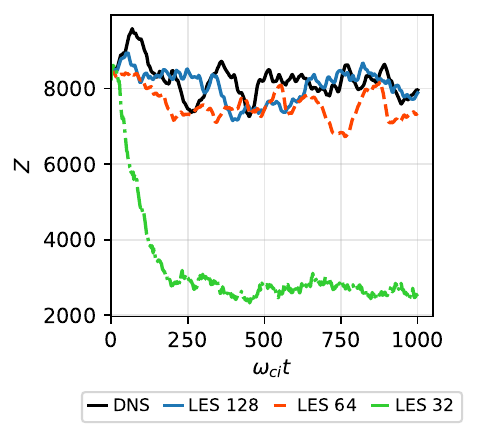}
\caption{HW\_N256}
\end{subfigure}
\hfill
\begin{subfigure}[t]{0.48\textwidth}
\centering
\includegraphics[width=\linewidth]{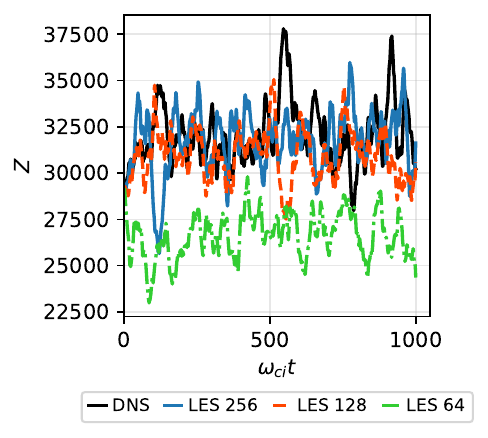}
\caption{HW\_N512}
\end{subfigure}
\par\smallskip
\begin{subfigure}[t]{0.48\textwidth}
\centering
\includegraphics[width=\linewidth]{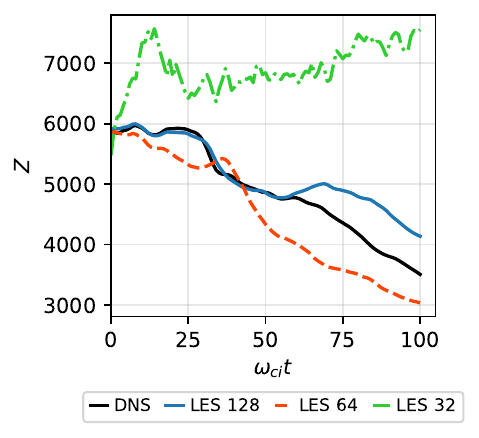}
\caption{mHW\_N256}
\end{subfigure}
\hfill
\begin{subfigure}[t]{0.48\textwidth}
\centering
\includegraphics[width=\linewidth]{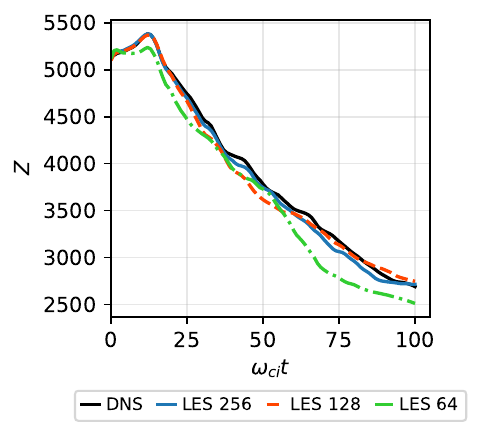}
\caption{mHW\_N512}
\end{subfigure}
\caption{Generalized enstrophy over $0\leq t\leq1000$ for HW and $0\leq t\leq100$ for mHW. The coarse HW reconstructions have reduced enstrophy; the $32^2$ mHW\_N256 reconstruction instead develops excess enstrophy, while mHW\_N512 shows decay after an initial peak.}
\label{fig:enstrophy}
\end{figure}

Figures~\ref{fig:energy} and~\ref{fig:enstrophy} reveal a strong resolution dependence over the longer HW interval. In HW\_N256, DNS energy fluctuates around $4.1$--$5.4\times10^3$. The $128^2$ and $64^2$ LES meshes retain comparable energy levels, although their fluctuations do not remain synchronized with DNS due to the chaotic nature of the flow. The $32^2$ LES instead decays rapidly to approximately $1.0$--$1.3\times10^3$ after $t\simeq150$. Its generalized enstrophy falls to roughly $2.5$--$3.0\times10^3$, compared with DNS levels of approximately $7.5$--$9.5\times10^3$. Thus the coarse reconstruction loses consistency in both metrics over the longer interval.

For HW\_N512, DNS energy is typically $1.4$--$1.7\times10^4$, with intermittent larger fluctuations. The $256^2$ LES retains this range, while $128^2$ generally lies lower and $64^2$ remains near $1.1$--$1.4\times10^4$. Enstrophy shows a similar ordering: the coarsest curve is mostly around $2.3$--$2.9\times10^4$, below the DNS range of roughly $2.9$--$3.8\times10^4$. We observe good agreement in the fluctuation levels and statistical behaviour, rather than pointwise comparison the turbulent trajectory.

For mHW\_N256, DNS energy rises from approximately $3.55\times10^3$ to $4.33\times10^3$ near $t=30$, before declining to $3.86\times10^3$ at $t=100$. The $128^2$ and $64^2$ reconstructions end near $3.86\times10^3$ and $3.90\times10^3$, respectively. The $32^2$ reconstruction instead continues growing to $5.23\times10^3$. Its generalized enstrophy reaches $8.81\times10^3$, compared with $3.51\times10^3$ for DNS and $3.54\times10^3$ and $3.76\times10^3$ for the two finer meshes. Thus the coarsest reconstruction fails to reproduce the decay and develops excess enstrophy well before the end of the mHW validation window.

In mHW\_N512, DNS energy initially rises from approximately $3.3\times10^3$ to $3.6\times10^3$ near $t=12$--$15$, then falls to about $3.0\times10^3$ at $t=100$. Generalized enstrophy similarly peaks near $5.4\times10^3$ before declining to about $2.7\times10^3$. The $256^2$ LES closely follows both trends. The $128^2$ and $64^2$ energy curves end near $2.9\times10^3$ and $2.75\times10^3$, respectively, with corresponding enstrophy deficits. This decay is consistent with the progression toward laminarization motivating the shorter mHW validation window.

\subsection{Spectral statistics}

The radial shell spectra include density and potential-gradient energy,
\begin{equation}
 E(k_s)=\frac12\sum_{\mathbf{k}\in\mathcal{S}_s}
 \left(|\widehat n_{\mathbf{k}}|^2+|\mathbf{k}|^2|\widehat\phi_{\mathbf{k}}|^2\right),
\end{equation}
where Fourier coefficients are normalized by the number of grid points and $\mathcal{S}_s$ is a radial wavenumber shell. This is a spectral energy density convention; summing the unsmoothed shells gives energy per unit area with spectral derivatives, rather than the finite-difference spatial integral above.

\begin{figure}[p]
\centering
\begin{subfigure}[t]{0.48\textwidth}
\centering
\includegraphics{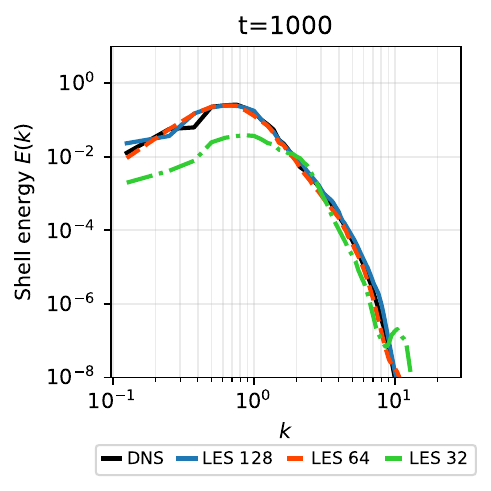}
\caption{HW\_N256}
\end{subfigure}
\hfill
\begin{subfigure}[t]{0.48\textwidth}
\centering
\includegraphics{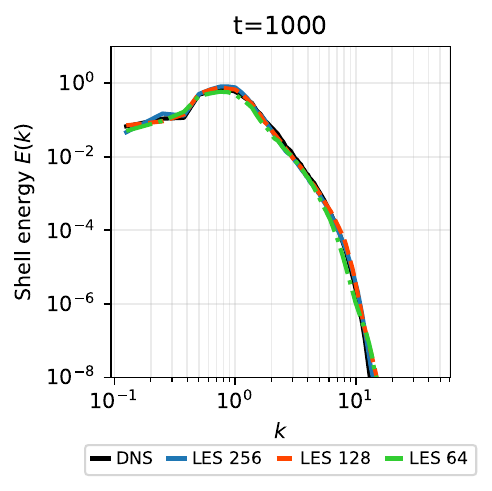}
\caption{HW\_N512}
\end{subfigure}
\par\smallskip
\begin{subfigure}[t]{0.48\textwidth}
\centering
\includegraphics{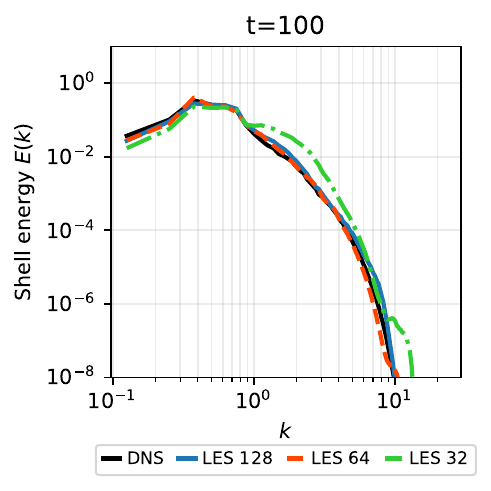}
\caption{mHW\_N256}
\end{subfigure}
\hfill
\begin{subfigure}[t]{0.48\textwidth}
\centering
\includegraphics{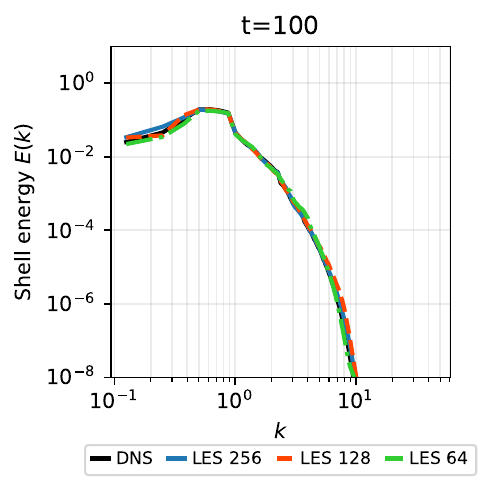}
\caption{mHW\_N512}
\end{subfigure}
\caption{Final ($t=1000$ for HW and $t=100$ for mHW) shell-summed HW energy spectra for the available cases and all three LES resolutions.}
\label{fig:spectra}
\end{figure}

Figure~\ref{fig:spectra} compares the final spectra. At $t=1000$, HW\_N256 with a $32^2$ LES mesh has a pronounced deficit across low and intermediate wavenumbers, consistent with its reduced integrated energy. A weak high-wavenumber tail remains, but does not compensate for the loss of energetic scales. The $128^2$ and $64^2$ LES spectra retain the main peak and broad decay more closely. HW\_N512 preserves the broad spectral shape across all three LES meshes, with differences at low wavenumbers, around the peak and in the weak dissipative tail. At $t=100$, the mHW\_N512 spectra remain close through the principal energy-containing range, with smaller differences toward the tail.

For mHW\_N256, the two finer LES meshes retain the final DNS spectral shape closely. The $32^2$ reconstruction has excess energy at intermediate and high wavenumbers, particularly for $k\gtrsim1$, consistent with its growing generalized enstrophy.

% Spectral agreement on logarithmic axes should be assessed together with the integrated diagnostics: similar shapes can coexist with energy deficits. The dotted line marks the $64^2$-grid Nyquist wavenumber in all panels; it is a fixed reference, not the cutoff of every LES mesh or a measured accuracy threshold.

\subsection{Probability-density functions}

\begin{figure}[p]
\centering
\begin{subfigure}[t]{0.48\textwidth}
\centering
\includegraphics{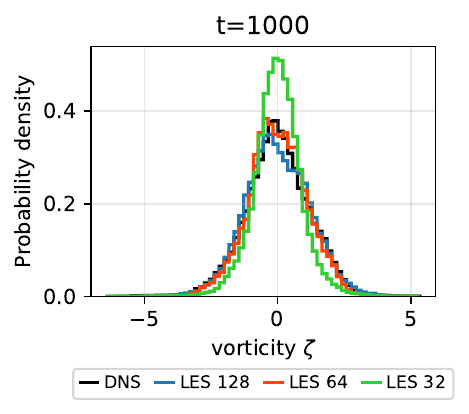}
\caption{HW\_N256}
\end{subfigure}
\hfill
\begin{subfigure}[t]{0.48\textwidth}
\centering
\includegraphics{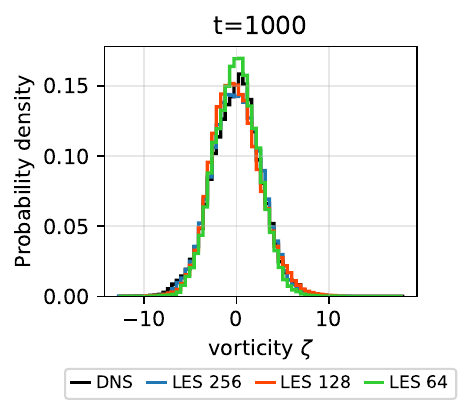}
\caption{HW\_N512}
\end{subfigure}
\par\smallskip
\begin{subfigure}[t]{0.48\textwidth}
\centering
\includegraphics{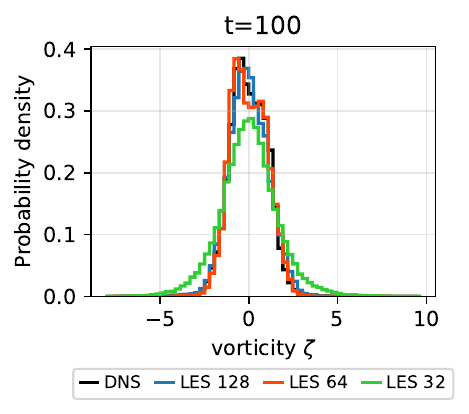}
\caption{mHW\_N256}
\end{subfigure}
\hfill
\begin{subfigure}[t]{0.48\textwidth}
\centering
\includegraphics{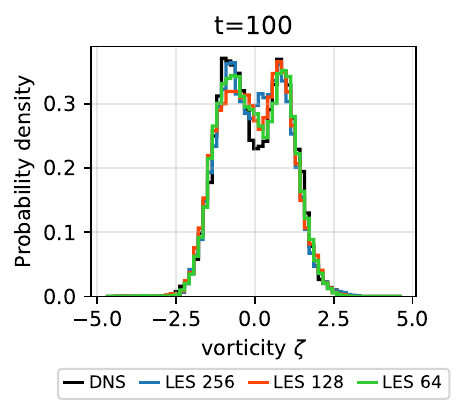}
\caption{mHW\_N512}
\end{subfigure}
\caption{Spatial PDFs of vorticity at $t=1000$ for HW or $t=100$ for mHW. Each available case includes DNS and three LES resolutions.}
\label{fig:pdf-vorticity}
\end{figure}

Figures~\ref{fig:pdf-vorticity} compare spatial PDFs for the vorticity at the final time of each model. These one-point statistics complement the field comparisons but do not measure spatial phase agreement or density--potential correlations. Their agreement can therefore coexist with differences in transport. The displayed linear-scale PDFs also do not establish quantitative accuracy for rare-event tails.

At $t=1000$, the HW\_N256 $32^2$ LES vorticity PDF becomes more sharply peaked. These changes agree with the loss of field amplitude indicated by energy and enstrophy. The two finer LES meshes retain broader distributions closer to DNS. For HW\_N512, the final distributions remain broadly similar, although the coarsest mesh sharpens the central vorticity peak.

At $t=100$, mHW\_N512 a broad structured with two-peaked vorticity PDF occurs in the DNS. The LES curves reproduce these qualitative shapes. Vorticity distributions differ locally around their peaks and the intervening troughs. Agreement of these spatial distributions does not imply that the instantaneous fields remain phase-aligned and should be investigated further with spatial comparisons and transport diagnostics.

In mHW\_N256, the $128^2$ and $64^2$ reconstructions retain the final vorticity PDFs more closely than $32^2$. At $t=100$, the coarsest vorticity PDF likewise broadens like the DNS, with a peak near $0.26$ versus $0.39$ for DNS. These changes support the energy and enstrophy evidence of excessive fluctuations on the coarsest mesh.

\subsection{Transport diagnostics}\label{sec:TransportDiagnostics}

The flux diagnostic removes the poloidal mean density, $n'=n-\langle n\rangle_y$, and uses $v_x=-\partial_y\phi$ and $v_y=\partial_x\phi$. The displayed domain averages are
\begin{equation}
 \overline\Gamma_x=-\langle n'\partial_y\phi\rangle_{x,y},\qquad
 \overline\Gamma_y=\langle n'\partial_x\phi\rangle_{x,y}.
\end{equation}
The same fourth-order periodic derivatives used for energy enter the flux calculation. A zonal velocity contributes no net flux to this diagnostic because the poloidal mean of $n'$ vanishes.

\begin{figure}[p]
\centering
\begin{subfigure}[t]{0.48\textwidth}
\centering
\includegraphics[width=\linewidth]{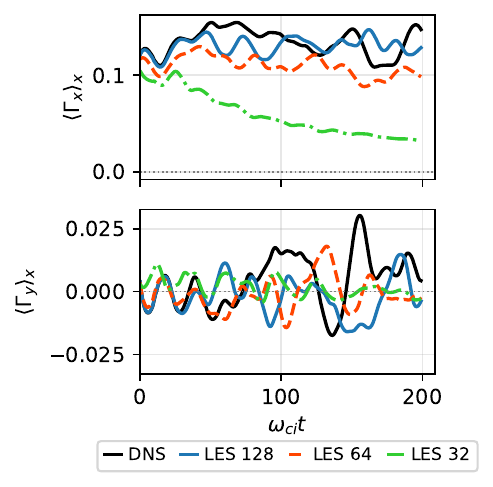}
\caption{HW\_N256}
\end{subfigure}
\hfill
\begin{subfigure}[t]{0.48\textwidth}
\centering
\includegraphics[width=\linewidth]{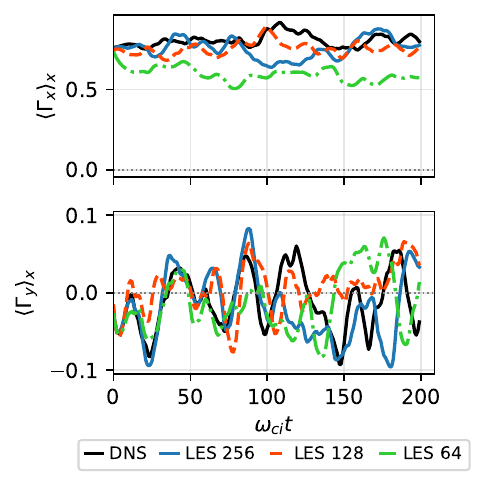}
\caption{HW\_N512}
\end{subfigure}
\par\smallskip
\begin{subfigure}[t]{0.48\textwidth}
\centering
\includegraphics[width=\linewidth]{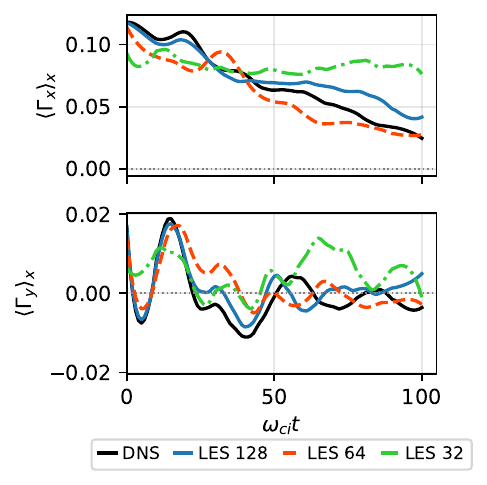}
\caption{mHW\_N256}
\end{subfigure}
\hfill
\begin{subfigure}[t]{0.48\textwidth}
\centering
\includegraphics[width=\linewidth]{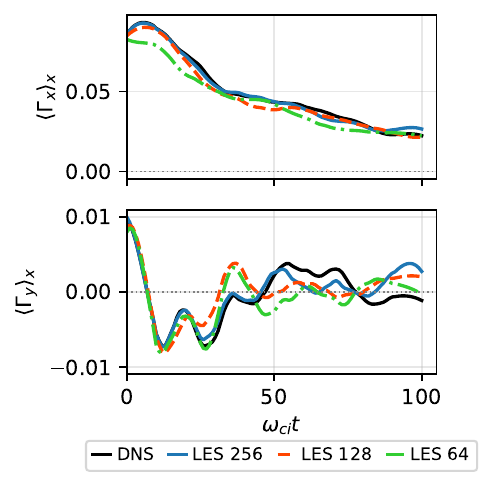}
\caption{mHW\_N512}
\end{subfigure}
\caption{Domain-averaged radial (upper axes) and poloidal (lower axes) turbulent particle flux over $0\leq t\leq1000$ for HW and $0\leq t\leq100$ for mHW. Density fluctuations are defined relative to the poloidal mean. All three LES resolutions are shown for each available case.}
\label{fig:particle-flux}
\end{figure}

Figure~\ref{fig:particle-flux} shows positive radial transport in all available cases. For HW\_N256, DNS fluctuates mainly around $0.11$--$0.15$, while the $32^2$ LES drops to approximately $0.04$ after the initial transient. The $128^2$ and $64^2$ meshes retain substantially more radial transport, although the latter is generally below DNS. In HW\_N512, DNS and the finest LES fluctuate mostly around $0.7$--$0.9$, whereas the coarsest mesh is typically around $0.55$--$0.65$. These deficits accompany the lower energy and enstrophy of the coarse solutions.

For mHW\_N256, DNS radial flux decreases from approximately $0.12$ to $0.026$ by $t=100$. The $128^2$ and $64^2$ reconstructions broadly follow this decline but end higher, at approximately $0.035$ and $0.033$. The $32^2$ reconstruction remains strongly fluctuating around $0.07$--$0.10$ after its initial transient and ends near $0.090$, more than three times the DNS value. Its excess transport accompanies the growth of energy and enstrophy rather than the DNS tendency toward laminarization. The finer meshes capture the early poloidal-flux extrema more closely; the coarsest mesh differs substantially in both timing and sign.

The mHW\_N512 radial flux peaks near $0.10$ early in the evolution and decreases to approximately $0.02$--$0.03$ by $t=100$. The finest LES follows the overall decline closely, while coarser meshes generally underpredict the early and intermediate transport. The marked decrease in transport accompanies the decay of generalized enstrophy as the flow evolves toward laminarization.

Poloidal transport fluctuates about zero and is smaller than radial transport. In HW\_N256, the $32^2$ LES strongly suppresses these fluctuations after the transient; the two finer meshes retain larger excursions but do not reproduce every DNS event. In HW\_N512, all curves remain strongly time dependent, with differences in the timing and amplitude of individual excursions. The mHW\_N512 LES curves capture the early negative excursions near $t\simeq13$ and $t\simeq27$ qualitatively, but differ from DNS later, including in sign. These comparisons distinguish preservation of the overall transport trend from agreement of instantaneous fluxes.

\subsection{Computational performance}

Wall-clock runtimes and MPI task counts were extracted from the final \texttt{Run time} entry and the \texttt{Processor number} entry, respectively, in the HW and mHW \texttt{BOUT.log.0} files. Each LES run is compared with the DNS run for the same model and reference resolution. The timing logs cover 1000 simulation time units for all 16 completed runs. These benchmark durations are distinct from the mHW statistical validation window, which is restricted to the first 100 time units. DNS uses 32 MPI tasks run on a Intel Xeon Gold 6448Y CPU 32-core, while LES uses 1 MPI task on 1-core of the same CPU attached to 1 GPU NVIDIA H100. Both CPU and GPU are from the
Mary Coombs supercomputer at the Hartree Centre (UK). 

We define:
\begin{equation}
    S_{\mathrm{wall}}=
    \frac{T_{\mathrm{DNS}}}{T_{\mathrm{LES}}},
    \qquad
    S_{\mathrm{MPI}}=S_{\mathrm{wall}}
    \frac{P_{\mathrm{DNS}}}{P_{\mathrm{LES}}},
\end{equation}
where $T$ is wall-clock runtime and $P$ is the number of MPI tasks. Here $S_{\mathrm{MPI}}$ measures the reduction in MPI task-time per simulated time unit; it is not a measured serial speedup. Values greater than unity indicate an improvement. Duration normalization compares average costs over the recorded trajectories rather than timings over an identical simulation interval.

\begin{table}[ht]
\centering
\small
\caption{Current HW and mHW runtimes extracted from \texttt{BOUT.log.0}. All runs cover 1000 simulation time units. Each DNS row is the baseline for its LES rows. Runtimes are displayed in minutes (m) and seconds (s), retaining the one-second precision reported by BOUT++. For each case, we highlight the best performing $S_\mathrm{wall}$ and $S_\mathrm{MPI}$ in bold.}
\label{tab:performance}
\begin{tabular}{lrrrr}
\toprule
Run & MPI tasks & Runtime & $S_{\mathrm{wall}}$ & $S_{\mathrm{MPI}}$ \\
\midrule
HW\_DNS256 & 32 & 4~m 36~s & \textbf{1.00} & 1.00 \\
HW\_DNS256\_LES32 & 1 & 7~m 19~s & 0.63 & \textbf{20.12} \\
HW\_DNS256\_LES64 & 1 & 12~m 42~s & 0.36 & 11.59 \\
HW\_DNS256\_LES128 & 1 & 27~m 26~s & 0.17 & 5.37 \\
\midrule
HW\_DNS512 & 32 & 59~m 3~s & 1.00 & 1.00 \\
HW\_DNS512\_LES64 & 1 & 42~m 25~s & \textbf{1.39} & \textbf{44.55} \\
HW\_DNS512\_LES128 & 1 & 92~m 5~s & 0.64 & 20.52 \\
HW\_DNS512\_LES256 & 1 & 320~m 41~s & 0.18 & 5.89 \\
\midrule
mHW\_DNS256 & 32 & 6~m 13~s & \textbf{1.00} & 1.00 \\
mHW\_DNS256\_LES32 & 1 & 21~m 27~s & 0.29 & 9.27 \\
mHW\_DNS256\_LES64 & 1 & 17~m 50~s & 0.35 & \textbf{11.16} \\
mHW\_DNS256\_LES128 & 1 & 41~m 17~s & 0.15 & 4.82 \\
\midrule
mHW\_DNS512 & 32 & 42~m 42~s & 1.00 & 1.00 \\
mHW\_DNS512\_LES64 & 1 & 27~m 38~s & \textbf{1.55} & \textbf{49.45} \\
mHW\_DNS512\_LES128 & 1 & 63~m 1~s & 0.68 & 21.68 \\
mHW\_DNS512\_LES256 & 1 & 185~m 37~s & 0.23 & 7.36 \\
\bottomrule
\end{tabular}
\end{table}

Table~\ref{tab:performance} shows MPI task-time reductions of \textbf{4.82--49.45} for the LES runs. The HW and mHW $512^2$ cases with $64^2$ LES meshes achieve wall-clock speedups of \textbf{1.39} and \textbf{1.55}, respectively. All other LES runs have $S_{\mathrm{wall}}<1$. These measurements use integer-second runtimes and cover 1000 simulation time units per run.

\section{Conclusions}\label{sec:Conclusions}

A hybrid AI-classical algorithm for edge-plasma LES closure has been developed by learning DNS fields filtered with a differential filter. We use the AI model to enhance the LES field towards a filtered DNS by adding details lost during the subsampling from DNS to LES resolutions. The DNS fields are then reconstructed by inverting the differential filter of \citet{germano_differential_1986}. The reconstructed fields are used to evaluate the nonlinear Poisson-bracket terms, while the vorticity is obtained directly from the generated electrostatic potential. The method is also applicable to other solvers in edge plasma physics and general purpose fluid dynamics solvers.

We benchmarked on the 2D Hasegawa-Wakatani and modified Hasegawa-Wakatani turbulent plasma models at $256^2$ and $512^2$ DNS resolutions using 2, 4 and 8 times smaller LES grids in each direction. Reconstructing from $32^2$ to $256^2$ resulted in poor prediction of the dynamics, as too much of the inertial subrange had been removed at this LES resolution, making the AI enhancement and differential filter inversion problematic. In all other cases, the time evolution statistics agreed well with the DNS for long time rollouts, showing enhanced stability compared to our previous study. Moreover, we showed the superiority of our approach to traditional algebraic LES models, which fail to capture most of the important dynamics of the turbulent flow. 

Overall the approach gave reasonable speedups of up to $1.55 \times$ on a single GPU (with LES on a single CPU), compared to DNS on 32 CPU cores. We expect a more significant speedup on larger 2D and future 3D studies. This will allow for accelerating not only edge plasma physics simulations for integration in powerplant design systems, but also several other applications in fluid dynamics.

% \begin{itemize}
%     \item the generative model reconstructs the dominant spatial structures of the DNS fields;
%     \item the finer HW reconstructions retain several statistical features over 1000 time units, while the $32^2$ LES for HW\_N256 substantially underpredicts energy, enstrophy and radial transport;
%     \item the mHW\_N512 reconstruction captures the growth and subsequent decay through 100 time units; this shorter window reflects faster dynamics and the tendency toward laminarization;
%     \item the thousand-time-unit HW and mHW benchmarks show MPI task-time reductions of 4.82--49.45, with a maximum wall-clock speedup of 1.55 for mHW\_DNS512\_LES64;
%     \item the $32^2$ mHW\_N256 reconstruction develops excess energy, enstrophy and radial transport, whereas the finer meshes follow the DNS decay more closely; finite-window agreement does not establish stability beyond the tested conditions.
% \end{itemize}

% Future work will investigate parameter-conditioned generators, explicit enforcement of conservation laws, uncertainty quantification and extension to fully three-dimensional edge-plasma models such as Hermes-3.
%----------------------------------------------------------------
%
%
%
%
%
%
%
%----------------------------------------------------------------
\section*{Acknowledgements}

Simulations reported in this study were performed on the Mary Coombs Hartree supercomputer platform. The authors acknowledge funding from the Fusion Computing Lab collaboration between the Hartree Centre and UKAEA and from the UK-USA AI Partnership Programme (ISPF-240).

\bibliographystyle{elsarticle-harv}
\bibliography{references}

\end{document}